# Dynamic Complex Network Analysis of $PM_{2.5}$ Concentrations in the UK using Hierarchical Directed Graphs

Broomandi, Parya[1,2,6]

Geng, Xueyu[1]

Guo, Weisi[3,1,4]

Kim, Jong[2]

Pagani Alessio[4]

Topping, David[5,4]

[1)] School of Engineering, The University of Warwick, Coventry, CV4 7AL, UK.

[2)]Department of Civil and Environmental Engineering, Nazarbayev University, 010000, Astana, Kazakhstan.

[3)]School of Aerospace, Transport, and Manufacturing, Cranfield University, Bedford, MK43 0AL, UK.

[4)]The Alan Turing Institute, London, UK.

[5)]School of Earth, Atmospheric and Environmental Science, University of Manchester, Manchester M13 9PL, UK.

[6)] Department of Chemical Engineering, Masjed-Soleiman Branch, Islamic Azad University, Masjed-Soleiman, Iran.

*Abstract*

Worldwide exposure to fine atmospheric particles can exasperate the risk of a wide range of heart and respiratory diseases, due to their ability to penetrate deep into the lungs and blood streams. Epidemiological studies in Europe and elsewhere have established the evidence base pointing to the important role of $PM_{2.5}$ (fine particles with a diameter of 2.5 microns or less) in causing over 4 million deaths per year. Traditional approaches to model atmospheric transportation of particles suffer from high dimensionality from both transport and chemical reaction processes, making multi-sale causal inference challenging. We apply alternative model reduction methods – a data-driven directed graph representation to infer spatial embeddedness and causal directionality. Using $PM_{2.5}$ concentrations in 14 UK cities over a 12-month period, we construct an undirected correlation and a directed Granger causality network. We show for both reduced-order cases, the UK is divided into two a northern and southern connected city communities, with greater spatial embedding in spring and summer. We go on to infer stability to disturbances via the network trophic coherence parameter, whereby we found that winter had the greatest vulnerability. As a result of our novel graph-based reduced modeling, we are able to represent high-dimensional knowledge into a causal inference and stability framework.

## 1. Introduction:

### 1.1 Background and rationale

Atmospheric particulate matter can be attributed to both local emissions (by both stationary and mobile sources) and regional transport processes. Causal inference between primary (emitted directly by the emission sources) and secondary (produced in the atmosphere by the transformation of gaseous pollutants) is challenging. For example, whilst combustion sources such as road traffic account for the bulk of anthropogenic PM emissions and cause $PM_{2.5}$ formation (Munir, 2017; AQEG, 2012), meteorological conditions can also influence $PM_{2.5}$ concentrations through dispersion, and deposition. Due to the high data complexity and dimensionality caused by the contribution of atmospheric chemistry transport processes and a range of emission sources in ambient $PM_{2.5}$ concentrations, we need to overcome the high dimensionality challenge and compress the concentration data into 2-dimensional (2D) network. European legislation sets current and future caps on anthropogenic emissions of primary and secondary-precursor components of $PM_{2.5}$ at national level and from individual sources (Vieno et al., 2016). In addition, it is well-known that ambient PM derives from both transboundary emissions and transport (Vieno et al., 2016), creating challenges to develop effective mitigation scenarios at the local level (Vieno et al., 2016; Zhang et al., 2008; van Donkelaar et al., 2010).

### 1.2 Importance & Impact

Atmospheric particulate matters impact human health (WHO, 2006, 2013) and climate change through radiative forcing (IPCC, 2013). The global health burden from exposure to ground level $PM_{2.5}$ is substantial. According to the Global Burden of Disease project, exposure to ambient $PM_{2.5}$ concentrations prevailing in 2005 was responsible for 3.2 million premature deaths and 76 million disability-adjusted life years (Vieno et al., 2016; Lim et al., 2012). In Europe, exposure to ambient $PM_{2.5}$ is still a major health issue. For the period 2010–2012, it was reported by the European Environment Agency report that 10–14 % of the urban population in the EU28 countries were exposed to $PM_{2.5}$ exceeding the EU annual-mean $PM_{2.5}$ reference value (25 $\mu g\ m^{-3}$), while 91–93 % were exposed to concentrations exceeding the WHO annual-mean $PM_{2.5}$ (10 $\mu g\ m^{-3}$) (Gehrig et al., 2003; EEA, 2014). Meeting the standards focused on $PM_{2.5}$ is complicated by the considerable chemical heterogeneity. PM long-term exposure has been identified to be more significant than the daily (short-term) exposure to higher levels of PM that had first been linked to health effects (Harrison et al., 2012; Pope and Dockery, 2006). Long-term impact studies have formed the basis for calculation of health outcomes from PM exposure in the UK and Europe, which are not insubstantial (COMEAP,2010). The re-orientation of attention towards $PM_{2.5}$, coupled with the evidence that long-term concentrations play important role alongside short-term peaks, in terms of health outcomes, has caused changes in legislation (Defra, 2007, Official Journal, 2008).

### 1.3 Modeling Challenges

Challenges associated with traditional modelling of PM evolution to infer regional and local influences include the need to embed a chemical complexity, range of emission sources and transformative processes in Eularian models. In this study, for the first time, we explore the potential for compressing ambient $PM_{2.5}$ network data into 2-dimensional (2D) network, establishing a simple graph to infer causality and stability. This is a timely study as strategic investments in national and local air quality monitoring networks require an evaluation on the usefulness, or not, of network design. Whilst this study focuses on a sparse distributed network, we discuss future applications for local networks across cities, for example. In a graph, each node in the graph is a city, which exhibits a temporal signal ($PM_{2.5}$) and is connected to other cities if they exhibit a close association in terms of either correlation (undirected) or Granger causality (directed).

## 2. Materials and Methods:

### 2.1 Ground-level $PM_{2.5}$ data

Hourly $PM_{2.5}$ concentrations were observed at 15 monitoring stations in different cities (from UK-air defra dataset website[1]) shown in Figure 1 and coordinates given in SI – List S1. The study period was divided into four seasons (meteorological seasons) Spring: 1st March 2017-31st May 2017, Summer: 1st June 2017- 31st August 2017, Autumn: 1st September 2017- 30th November 2017, and Winter: 1st December 2017- 28th February 2018. Also, $PM_{2.5}$ emissions sources data were downloaded from the UK National Atmospheric Emission Inventory (NAEI) website.

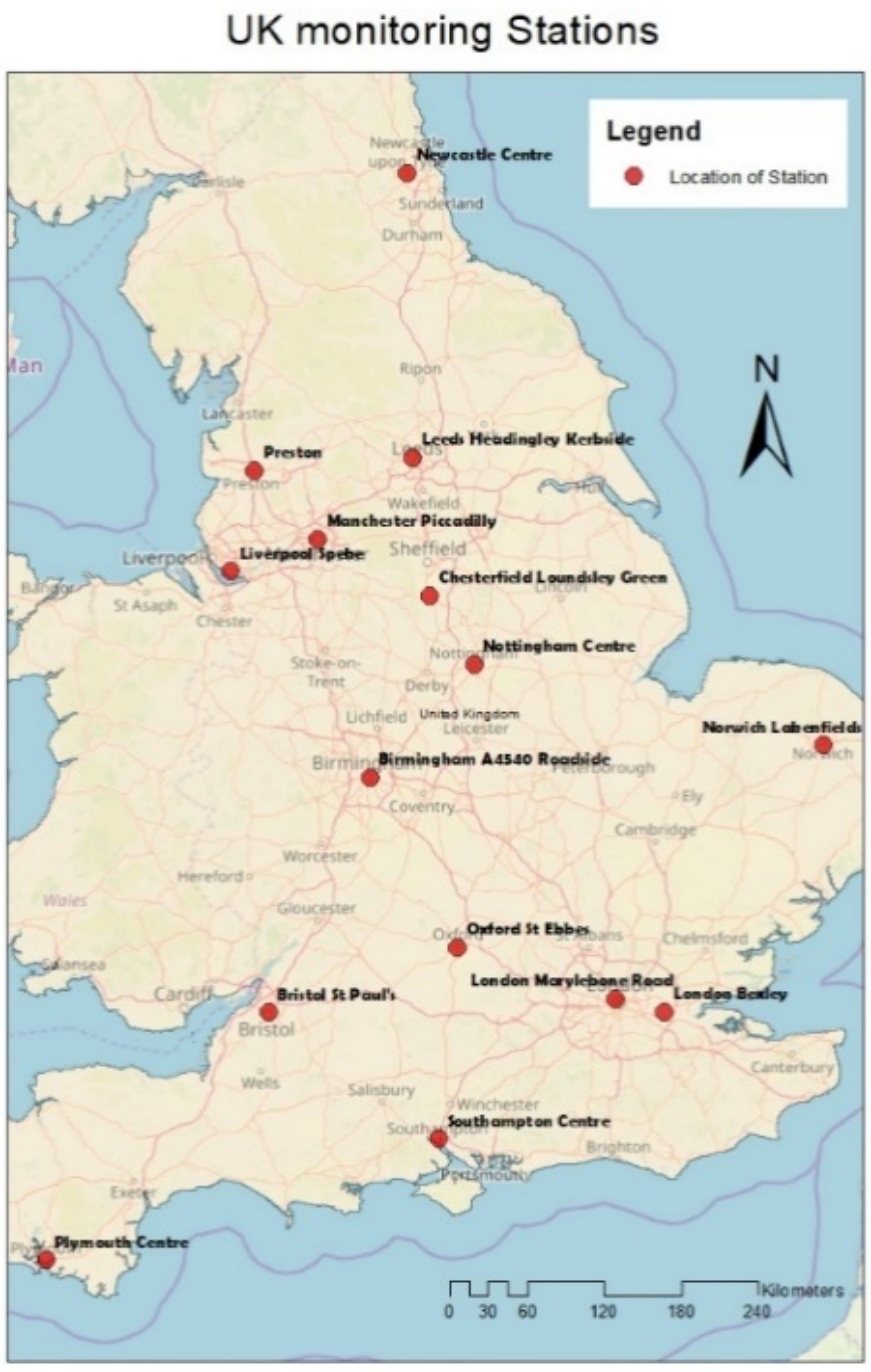

Figure 1. Studied stations in the UK.

## 2.2 Cross correlation calculation for spatial distribution of $PM_{2.5}$ in the UK

To measure the similarity of $PM_{2.5}$ concentration time series among each pair of cities in the current study, the hourly based cross-correlation (XCROSS) was calculated using PAST (PAleontological Statistics) version 3.25, for all site pairs (106 pair of cities) in four seasonal windows (spring, summer, autumn, and winter). These periods were selected to try and capture the effect of seasonal changes on the measured similarity between $PM_{2.5}$ concentration time series. A flexible threshold (above 70%) was applied to decide which pairs were strongly correlated (Gehrig et al., 2003).

[1] https://uk-air.defra.gov.uk/data/openair

## 2.3 Granger Causality calculation in $PM_{2.5}$ network in the UK

The Granger causality test as a statistical hypothesis test for determining whether one time series is useful in forecasting another, thus for measuring the ability to predict the future values of a time series using prior values of another time series, was applied (using Eviews, version 11) to each pair of cities in the network during different seasons. When the p-value was less than alpha level (5%), the null hypothesis was rejected, and we could decide which time series can forecast another one. The Granger Causality test assumes that both the $x$ and y time series (x and y represent $PM_{2.5}$ concentration series for different stations in our network) are stationary, which was not the case in current study. As a result, de-trending was first employed before using the Granger Causality test. To retain the same degrees of freedom (Statistical parameter estimation is based on different amounts of data or information. The number of independent pieces of data that go into the estimation of a parameter are called the degrees of freedom (DF). Mathematically, DF represents the number of dimensions of the domain of a random vector, or how many components should be known before the vector is fully determined.), with annual data, the lag number is typically small (1 or 2 lags). For quarterly data (which was our case), the appropriate lag number is 1 to 8. If monthly data is available, 6, 12, or 24 lags can be used given enough data points. The number of lags is critical since a different number of lags can lead to different test results. As a result, optimal lags were chosen based on Akaike Information Criterion (AIC). The optimal lag number that ensures the model will be stable is thus 7 in our study. It is possible that causation is only in one direction, or in both directions ($x$ Granger-causes y and y Granger causes $x$). We chose the direction based on the lowest p-value. For example in spring, according to our analysis, results suggest that ‘activity’ in Manchester is statistically influencing Preston with a p-value= $5\times10^{-29}$, while Preston is statistically affecting Manchester with a p-value= $3\times10^{-8}$. Therefore we infer that the first statement (pollution from Manchester is influencing Preston’s concentrations) is the correct one to select due to its lower p-value. Please note the language chosen reflects the statistical inference for the network analysis; However, the mapping of inference to atmospheric behavior and known challenges around PM2.5 source apportionment is important and discussed.

## 2.4 Trophic coherence

Trophic coherence is a way of hierarchically restructuring a directed network and labelling the hierarchical levels (trophic levels – as derived from food webs and predation levels). Trophic levels have been shown to be an effective compressed metric to infer stability on large directed networks with no clear input output definition. The bottom (basal) nodes are those where all energy comes from (e.g. major source of pollution), and the coherence of the whole network is a proxy for stability against disturbances. The trophic level ($s_i$) of a node i, is defined as the average trophic level of its in-neighbours:

$$s_i = 1 + \frac{1}{k_i^{in}} \sum_j a_{ij} s_j$$

where $a_{ij}$ is the adjacency matrix of the graph and $k_i^{in} = \sum_j a_{ij}$ is the number of in-neighbours of the node i. Basal nodes $k_i^{in}$ have trophic level $s_i = 1$ by convention (Pagani et al., 2019). In our study, to define trophic coherence in a directed causal network, the first step was defining basal nodes.

Stations with a low trophic level are $PM_{2.5}$ sources while stations with a high trophic level are receptors according to this definition. The trophic level of a station is the average level of all the stations from which it receives $PM_{2.5}$ pollutant plus 1. $x_{ij} = s_i - s_j$ is the associated trophic difference of each edge. As always, p(x) (the distribution of trophic differences) has a mean value of 1, and the more a network is trophically coherent, the smaller the variance of this distribution. The trophic coherence of network is measured with the incoherence parameter q, which is the standard deviation of p(x):

$$q = \sqrt{\frac{1}{L} \sum_{ij} a_{ij} x_{ij}^2 - 1},$$

where $L = \sum_{ij} a_{ij}$ is the number of connections (edges) between the stations (nodes) in the network. A perfectly coherent network has $q = 0$, but a q greater than 0 indicates less coherent networks.

## 3. Result and Discussion:

### 3.1 Spatial distribution of PM2.5 over the UK

Interesting information about the spatial distribution of the $PM_{2.5}$ concentrations over the UK can be obtained when analysing the cross correlation of the hourly values between the different sites. Results suggest that two groups of cities were connected to each other with XCROSS value above 70%. The first group (Northern Group A) includes Preston (Pre), Manchester (Man), Chesterfield (Chest), Leeds, Nottingham (Not), Newcastle (New), Birmingham (Bir), and Liverpool (Liv), while the second one (Southern Group B) includes Bristol (Bri), Oxford (Oxf), Southampton (South), Plymouth (Ply), Norwich (Nor), and London (two stations named LonB and LonR). For the seasons of spring, summer, and autumn, the combination of groups does not change, but the value of XCROSS does (Figure 2). In wintertime the combination of cities in and out of clusters changes (Figure 2-D). The connected cities, generating a directed dynamic network, are seasonally visualized in Figure 2.

As the networks are very spatial (i.e., distance is a significant impedance factor), a general measure of how spatially embedded it is, was studied. The pair of stations were divided into groups based on the distance (Table 1). To quantify the level of spatial embeddedness, a relationship between Cross correlation and distance between each pair of cities was studied (Table 1). A very high spatially embedded part of the network for all seasons was formed below 100 Km, while less spatial embeddedness of network was witnessed when the distance increased to above 200Km (for all seasons). A main part of the network (100 Km) was formed in cluster A with percentage of 67%, 54%, 60%, and 89% during spring, summer, autumn, and winter, respectively. This value in cluster A reduced (for all seasons) when increasing the distance between pair of cities reaching the value of zero during autumn and winter. Since the distance between cities in cluster was dominantly above 100Km, the dominant part of the network in cluster B was formed below 200 Km (100-200Km), with percentage of 38%, 52%, 46%, and 23% during spring, summer, autumn, and winter, respectively. This value in cluster B had a reduction (for all seasons) by increasing the distance between pair of cities reaching the value of zero during autumn, while during wintertime it was 19% for distance above 200Km. The number of outliers (pair of connected cities out of group A &B) had its highest values of 40%, 100%, and 81% during spring, autumn, and winter, respectively when the distance between cities was above 200Km. During autumn, for distances above 200Km, the original network was not formed, while during winter, group B was formed. The number of paired cities in the network had a reduction by 50% between spring and winter, when the distance was below 100Km (the same reducing trend was witnessed in both groups). For distances below 200Km, the network was weakened by %50. Interestingly, when the distance between cities increased above 200 Km, during winter the network was strengthened by 17% comparing to spring.

Table 1. The relationship between Cross-Correlation (XCROSS) of the daily values of $PM_{2.5}$ and distance of the cities in UK.

| Distance | Pair of connected cities in network | Pair of connected cities in group A | Pair of connected cities in group B | Outliers (pair of connected cities out of groups) |
|---|---|---|---|---|
| **Spring** | | | | |
| <100Km | 18 (43%) | 12 (67%) | 6 (33%) | 0 |
| <200Km | 42 (81%) | 24 (57%) | 16 (38%) | 2 (5%) |
| >200Km | 10 (19%) | 3 (30%) | 3 (3%) | 4 (40%) |
| **Summer** | | | | |
| <100Km | 13 (52%) | 7 (54%) | 6 (46%) | 0 |
| <200Km | 25 (90%) | 12 (48%) | 13 (52%) | 0 |
| >200Km | 3 (10%) | 2 (67%) | 1 (33%) | 0 |
| **Autumn** | | | | |
| <100Km | 15 (54%) | 9 (60%) | 6 (40%) | 0 |
| <200Km | 28 (93%) | 9 (27%) | 13 (46%) | 9 (27%) |
| >200Km | 2 (7%) | 0 | 0 | 2 (100%) |
| **Winter** | | | | |
| <100Km | 9 (35%) | 8 (89%) | 1 (11%) | 0 |
| <200Km | 26 (41%) | 14 (54%) | 6 (23%) | 6 (23%) |
| >200Km | 37 (59%) | 0 | 7 (19%) | 30 (81%) |

A

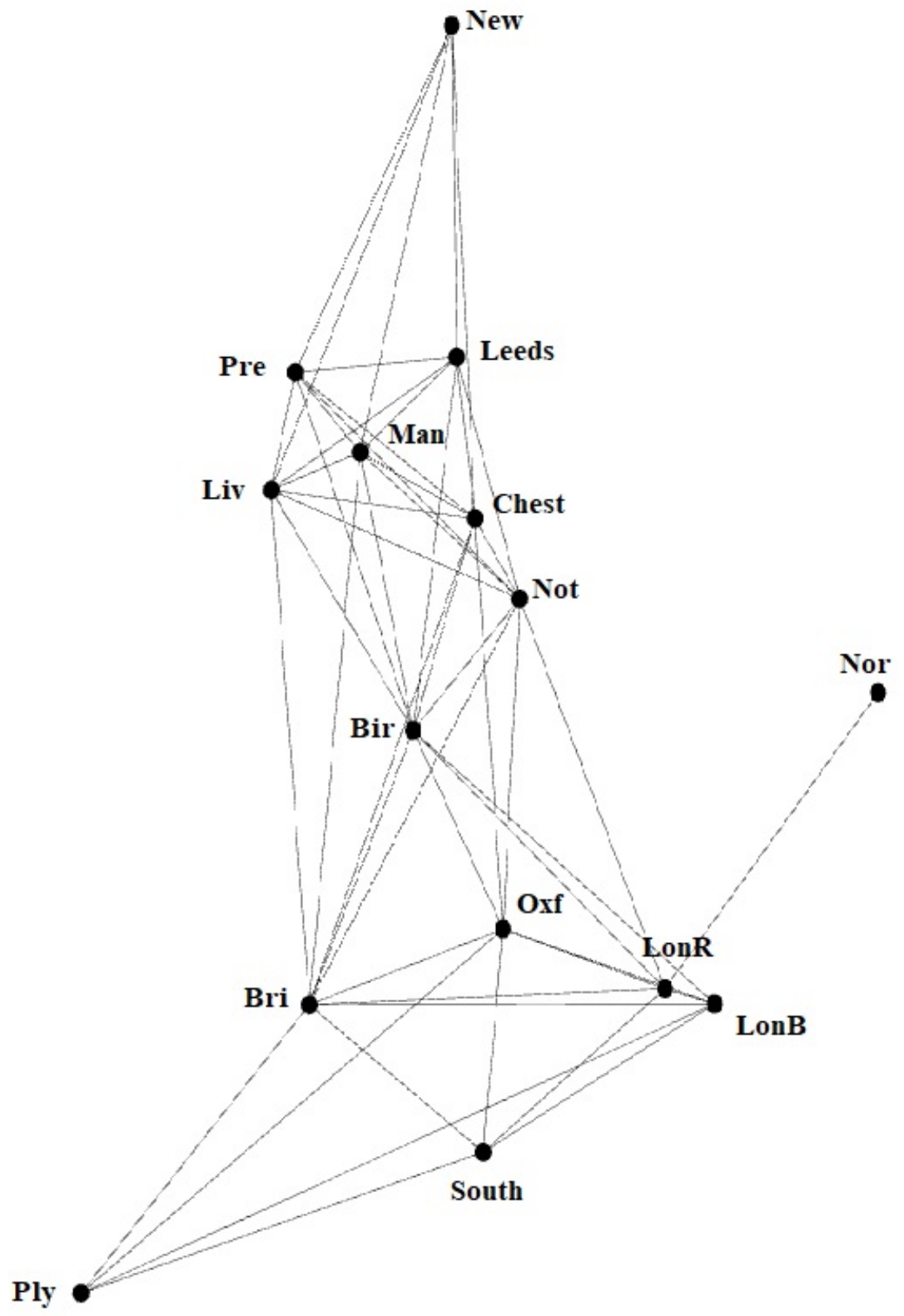

B

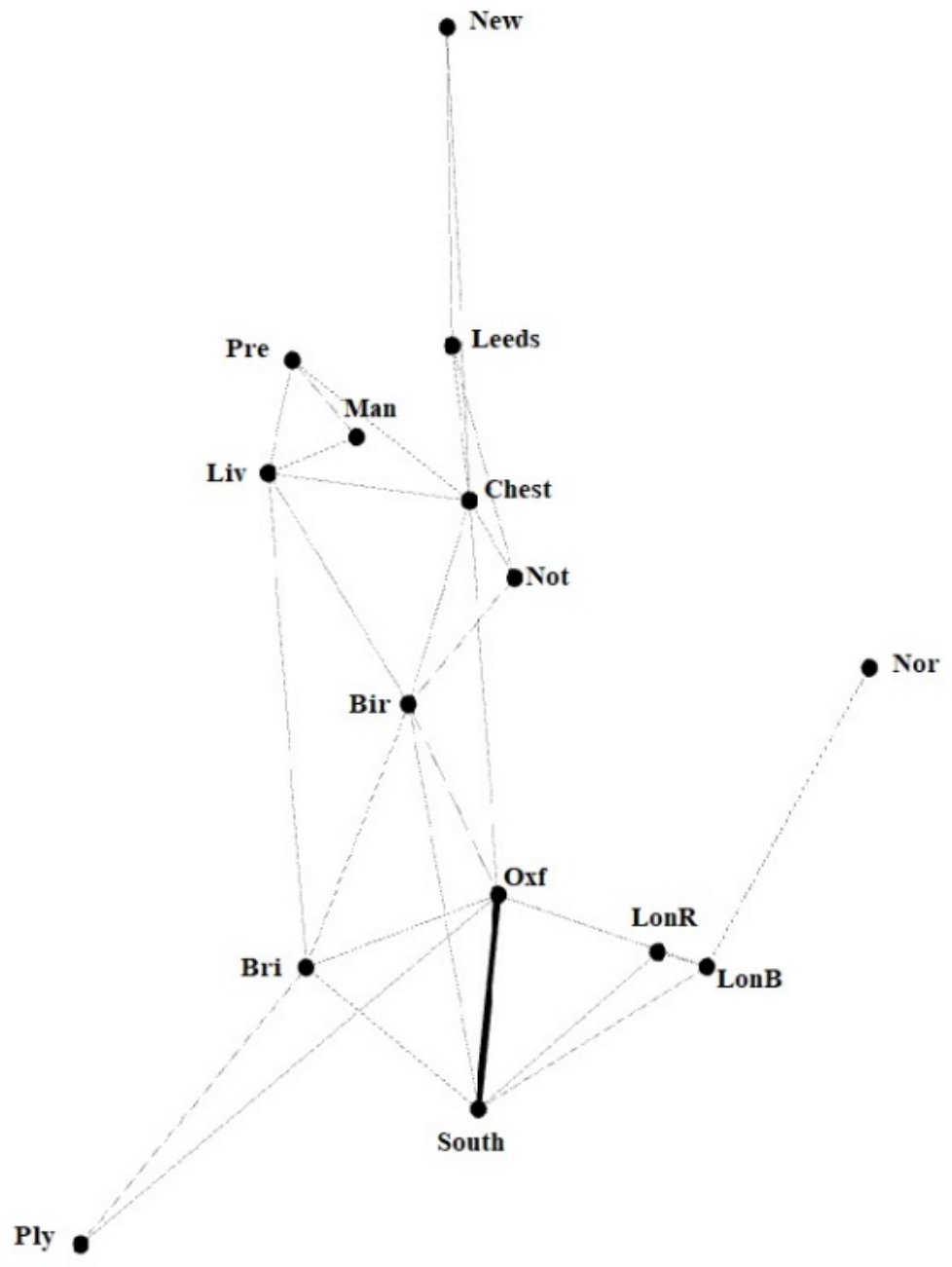

C

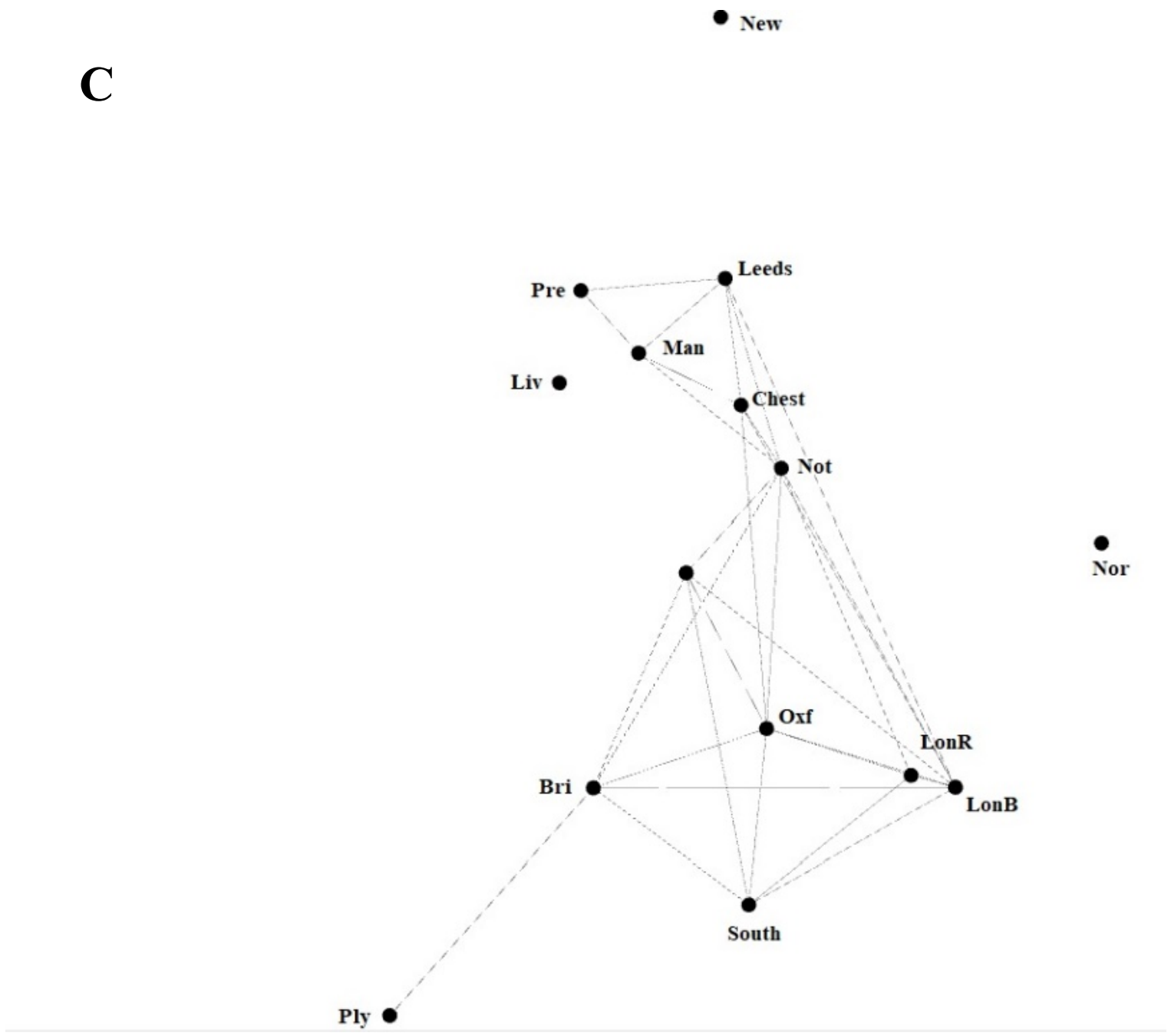

D

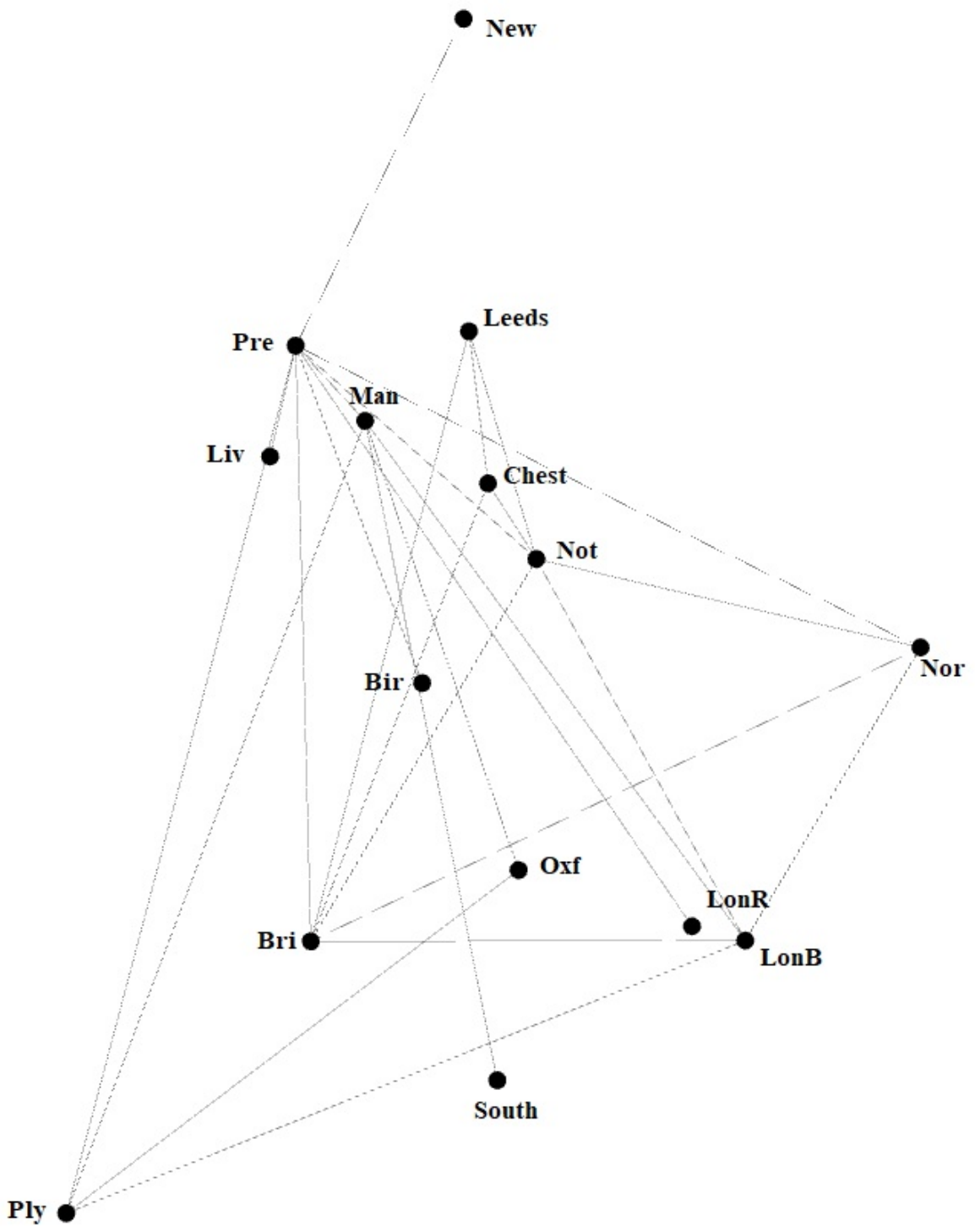

Figure 2. Cross correlation based dynamic network including; A) spring window, B) summer window, C) autumn window, and D) winter window in 2017-2018, UK.

3.2 Granger causality test

The main result from this study is that cities with the strongest Cross correlation have the lowest p-value (below 5%) (Figure 3). In spring, as already noted, results suggest that, statistically, activity in Manchester is causing concentrations to change in Preston with p-value= $5\times10^{-29}$ (i.e. Manchester $PM_{2.5}$ data can be used to predict the future $PM_{2.5}$ values of Preston) and Bristol is causing Oxford with a p-value of $9\times10^{-28}$. In summer, Liverpool is causing Preston with a p-value of $7\times10^{-17}$. Manchester is causing Preston with p-value= $6\times10^{-23}$ in autumn, while Chesterfield is causing Nottingham with a p-value of $1\times10^{-7}$in wintertime. The results look very spatial and the distance is a significant impedance factor. The distance between all paired cities was below 50Km. Based on Table 2, when the distance between pair of cities increases the order of p-value increases too.

**A**

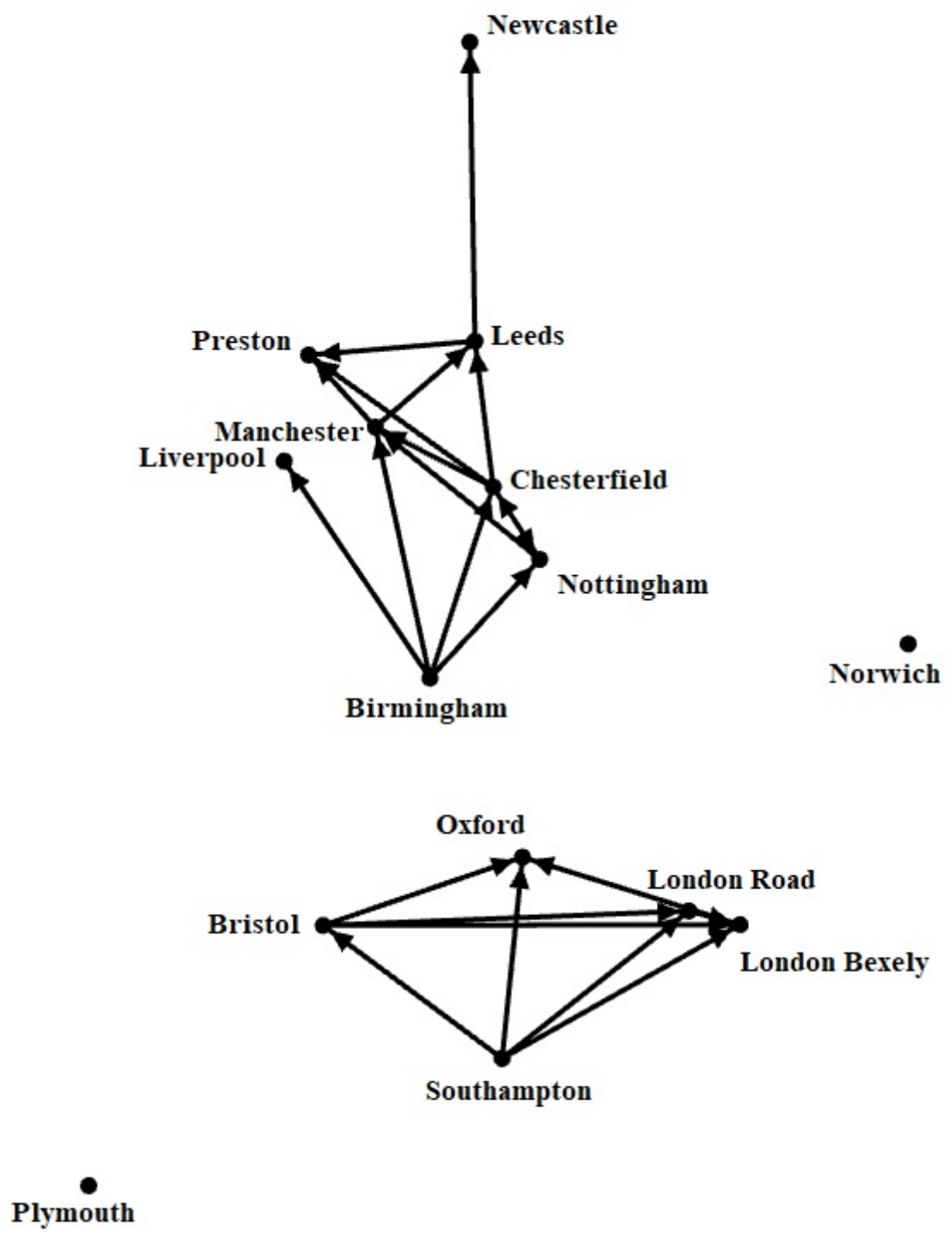

**B**

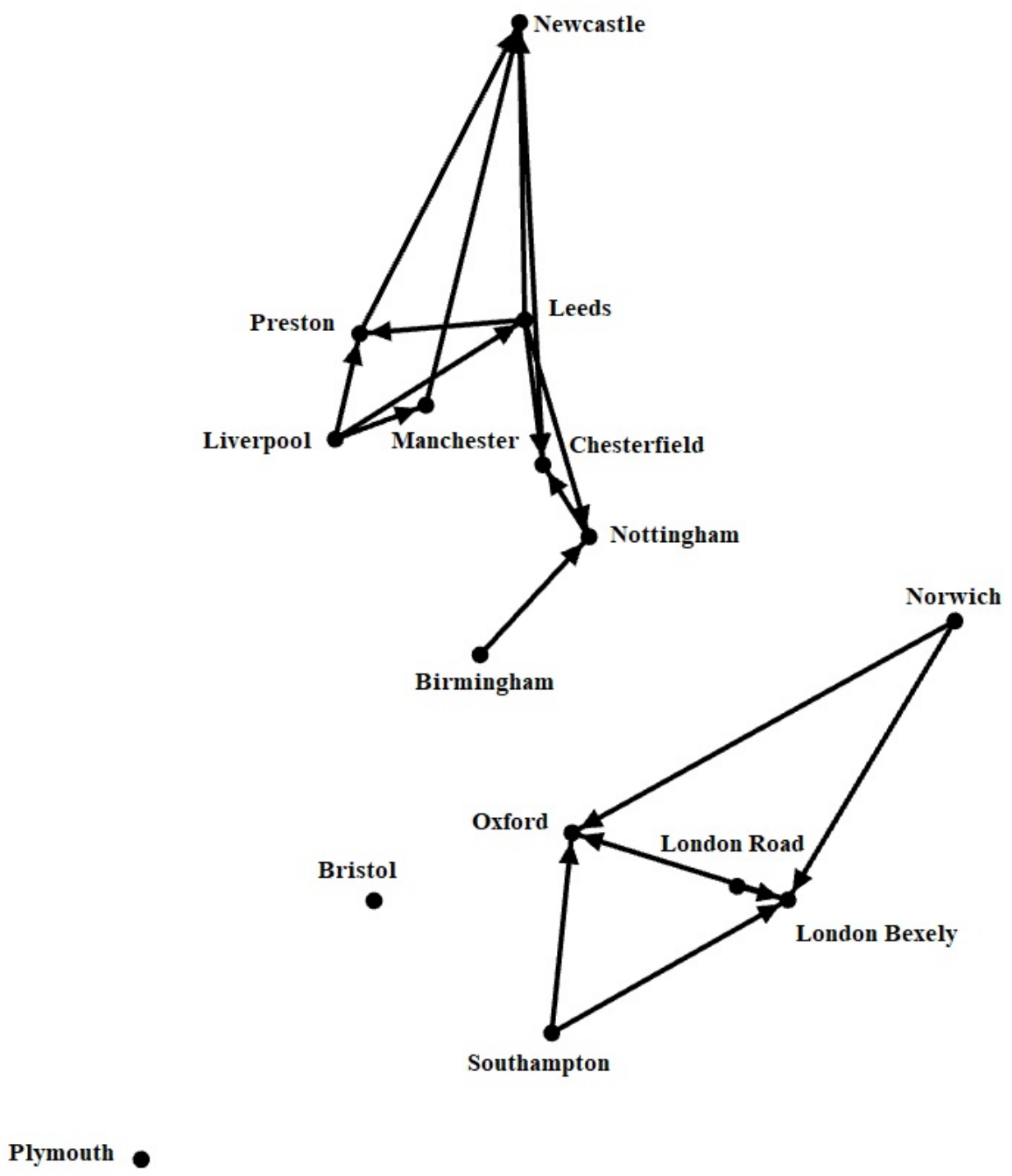

C

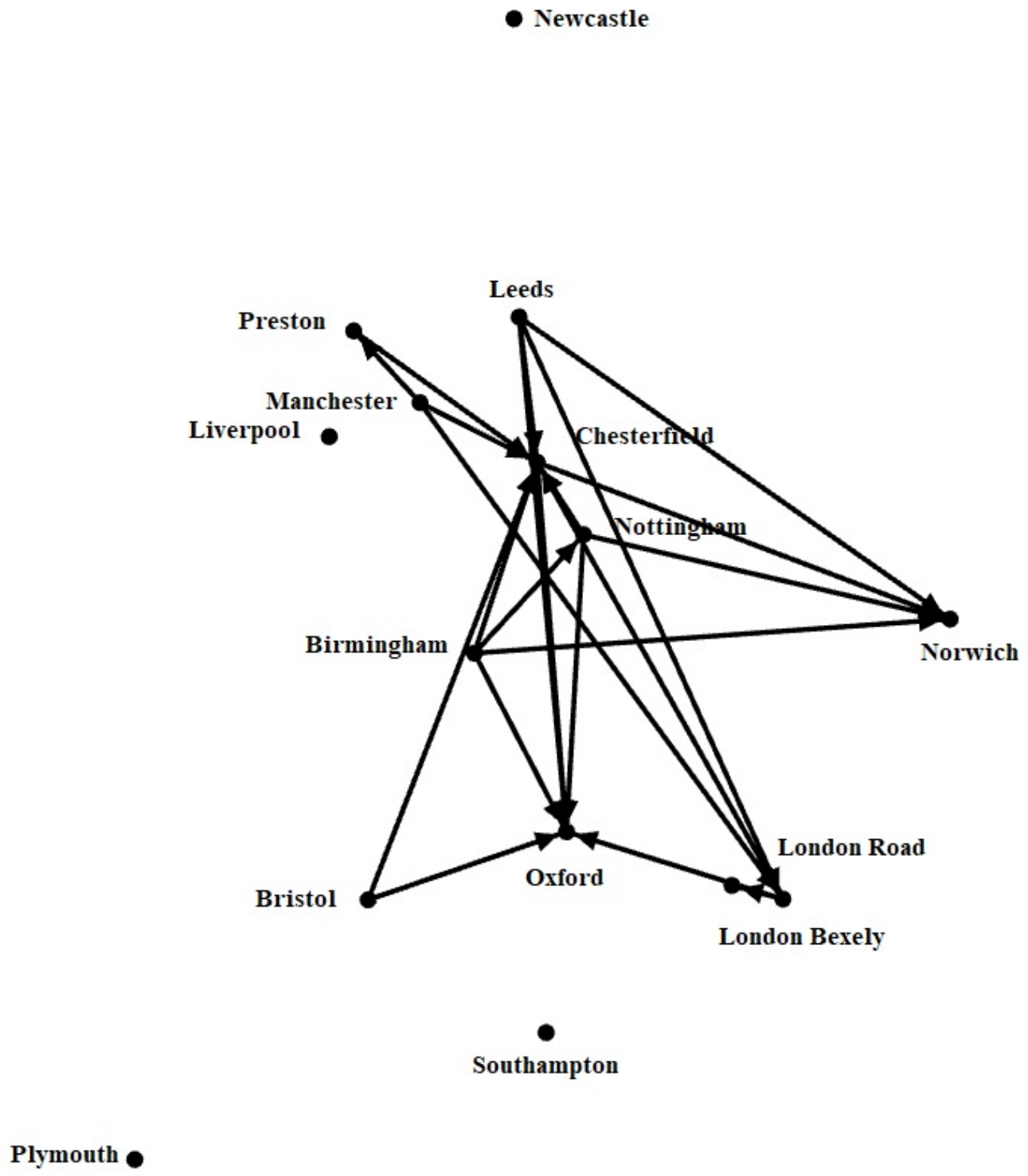

D

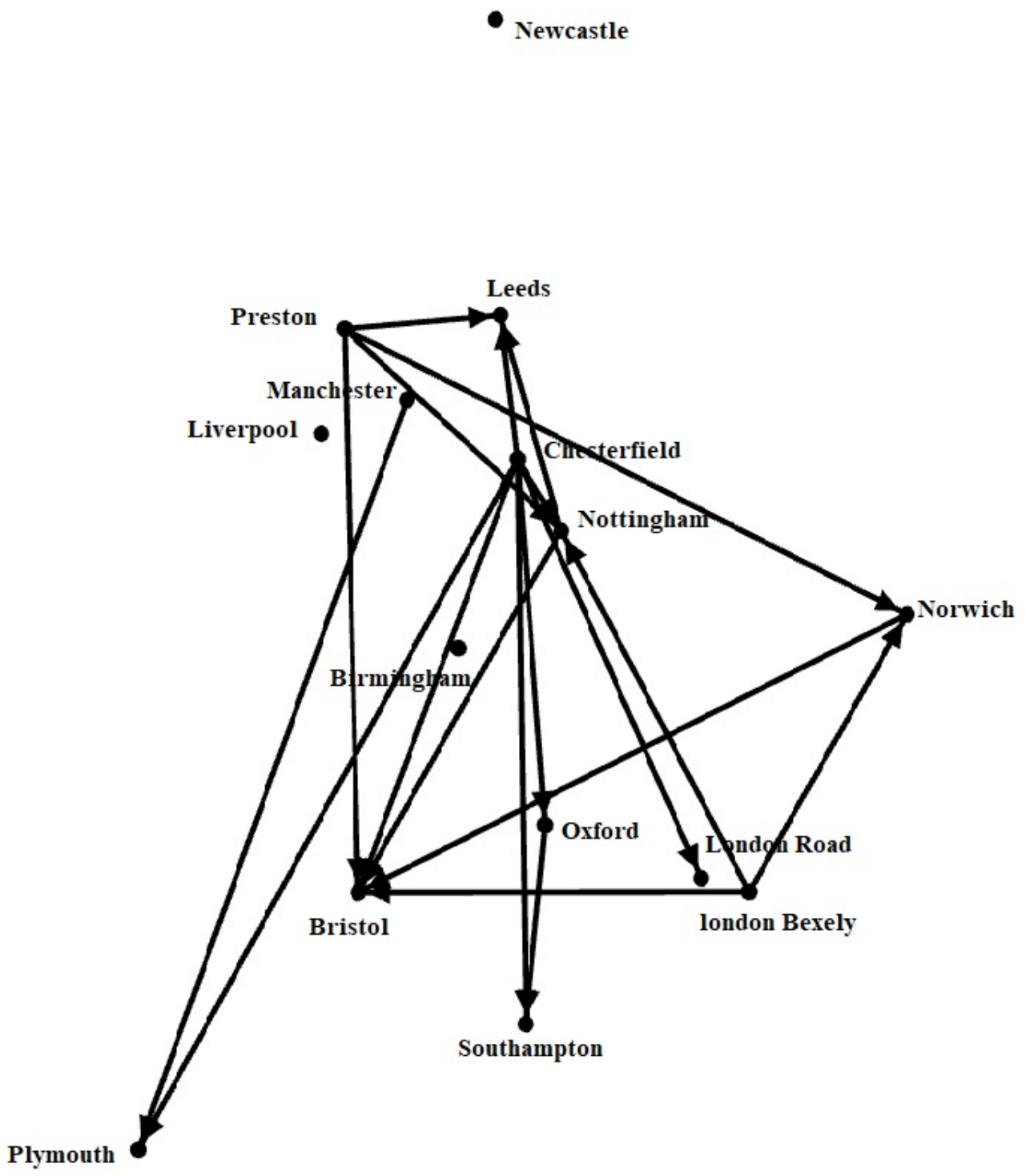

Figure 3. Granger based dynamic network including; A) Spring window, B) Summer window, C) Autumn window, and D) Winter window in 2017-2018, UK.

Table 2. Comparison among Granger causality results (p-values) in different seasons.

| **Source** | **Target** | **Distance** (Km) | **p-value** |
|---|---|---|---|
| | | **Spring** | |
| Manchester | Preston | 43.66 | $5\times10^{-29}$ |
| Bristol | Oxford | 91.78 | $9\times10^{-28}$ |
| | | **Summer** | |
| Liverpool | Preston | 42.62 | $7\times10^{-17}$ |
| Leeds | Newcastle | 131 | $5\times10^{-11}$ |
| | | **Autumn** | |
| Manchester | Preston | 43.66 | $6\times10^{-23}$ |
| Chesterfield | Oxford | 165.11 | $3\times10^{-20}$ |
| | | **Winter** | |
| Chesterfield | Nottingham | 36.17 | $1\times10^{-7}$ |
| Chesterfield | Bristol | 213.74 | $7\times10^{-6}$ |

A directed graph is defined (Bang-Jensen and Gutin, 2008) as an ordered pair $G=(N,E)$, where N is a set of nodes (i.e. stations) and E is a set of ordered pairs of nodes, called edges (i.e the probability values for F statistics). The hierarchical structure of a directed graph can be presented by its trophic coherence property. The whole idea is that hierarchical systems have fewer feedback loops and are less likely to suffer from cascade effects. We measured the coherence of the seasonal causal network through the incoherence parameter (q) as a measure of how tightly the trophic distance associated with edges is concentrated around its mean value (which is always 1) (Johnson, et al., 2014). We observed incoherent network in our seasonal datasets (Table 3).

Table 3. Incoherence factor of seasonal directed networks in current study.

| **Directed network** | **Incoherence factor (q)** |
|---|---|
| Spring | 0.69 |
| Summer | 0.37 |
| Autumn | 0.49 |
| Winter | 0.35 |

The highly incoherent season was spring with q= 0.69, whilst a less incoherent network was found to be winter (q=0.35). In figure 3, according to the parameter definition, the basal nodes with the low trophic level represent the major pollution source nodes, while stations with high trophic levels are ones who act as receptors in the causal network. During springtime, due to well mixing of the lower atmospheric layer, the network was well formed. In group A, Birmingham with low trophic level was classified as a pollution source, while in group B Southampton was pollution source with low trophic level.

**A**

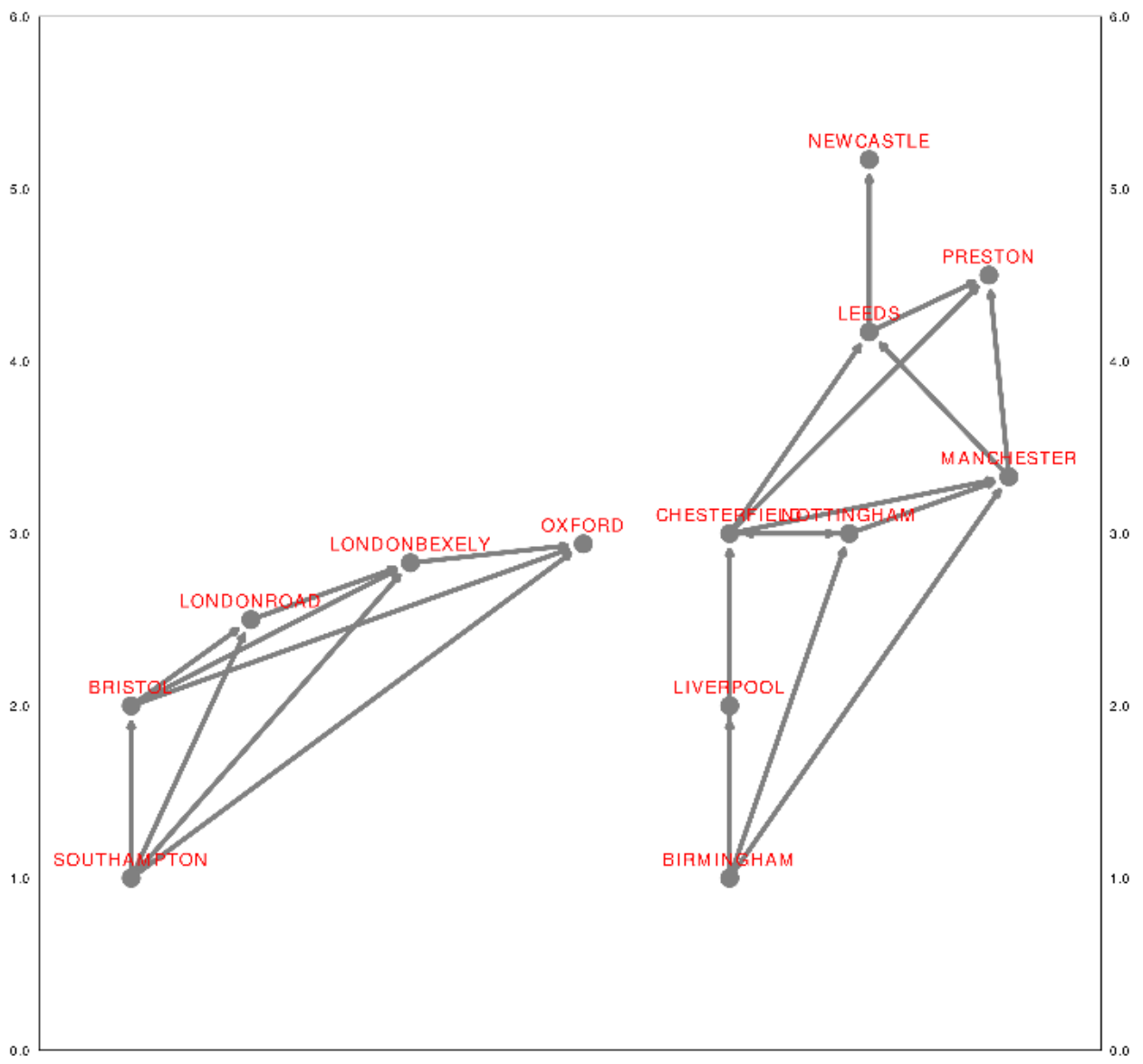

**B**

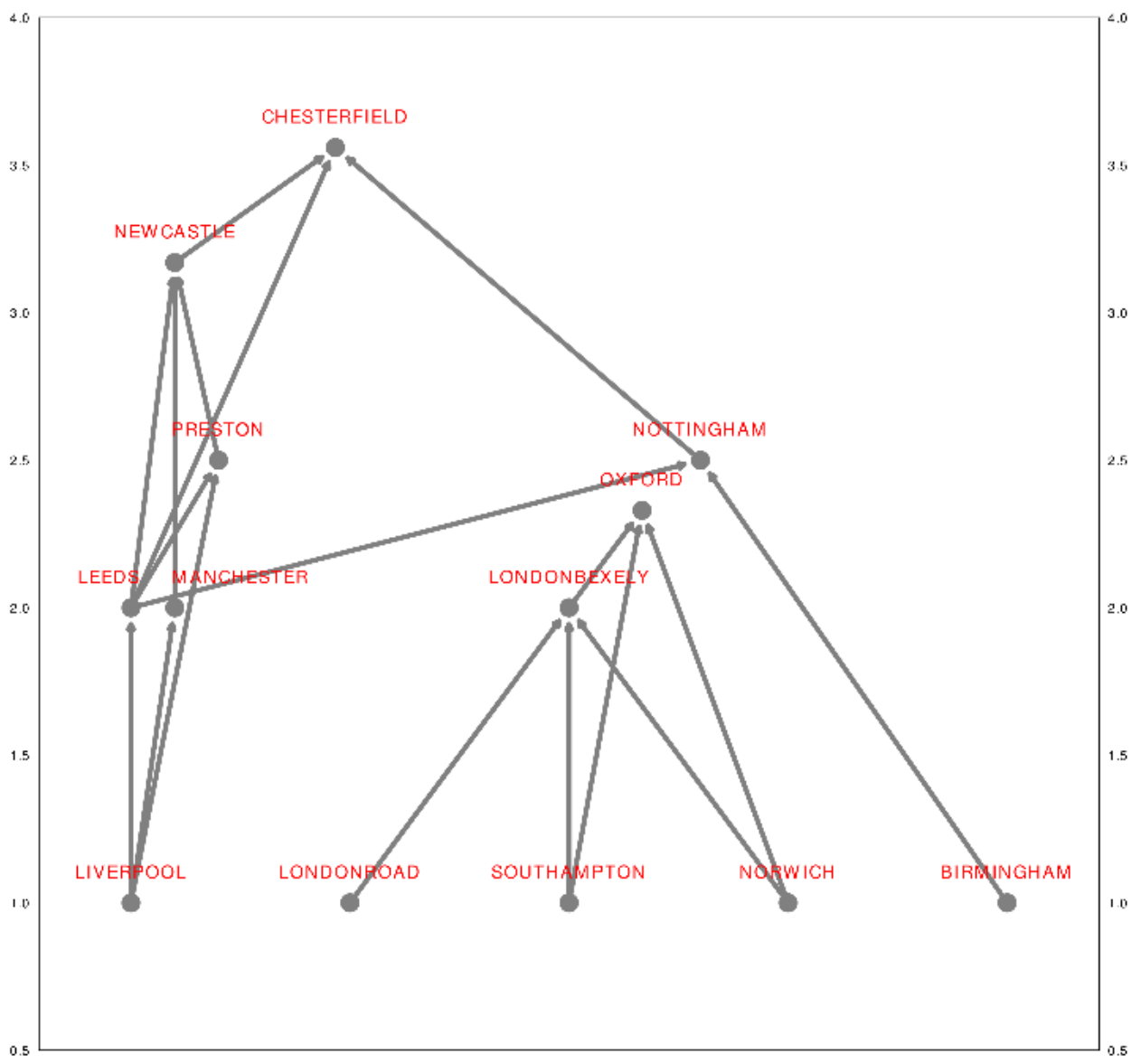

**C**

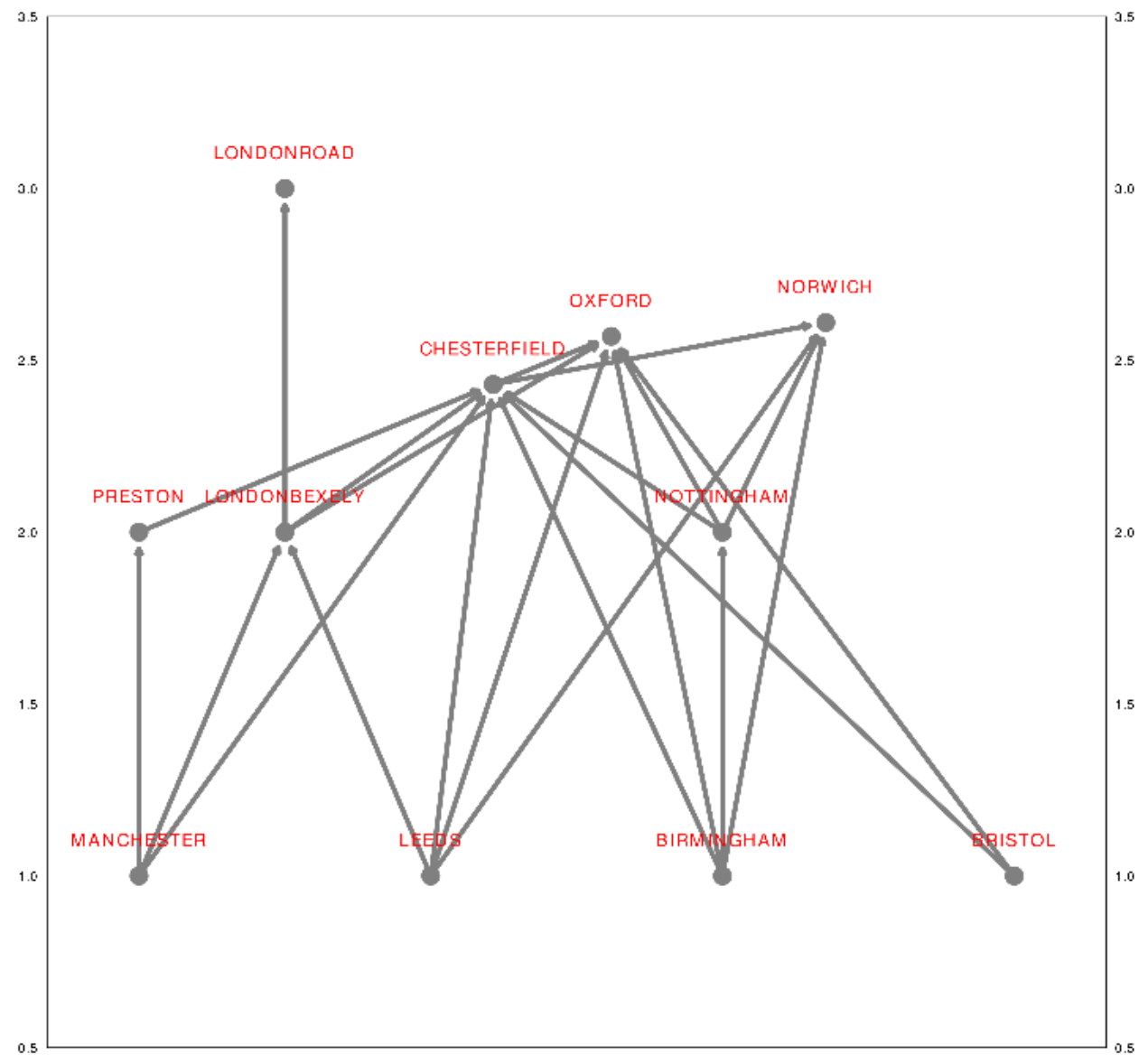

**D**

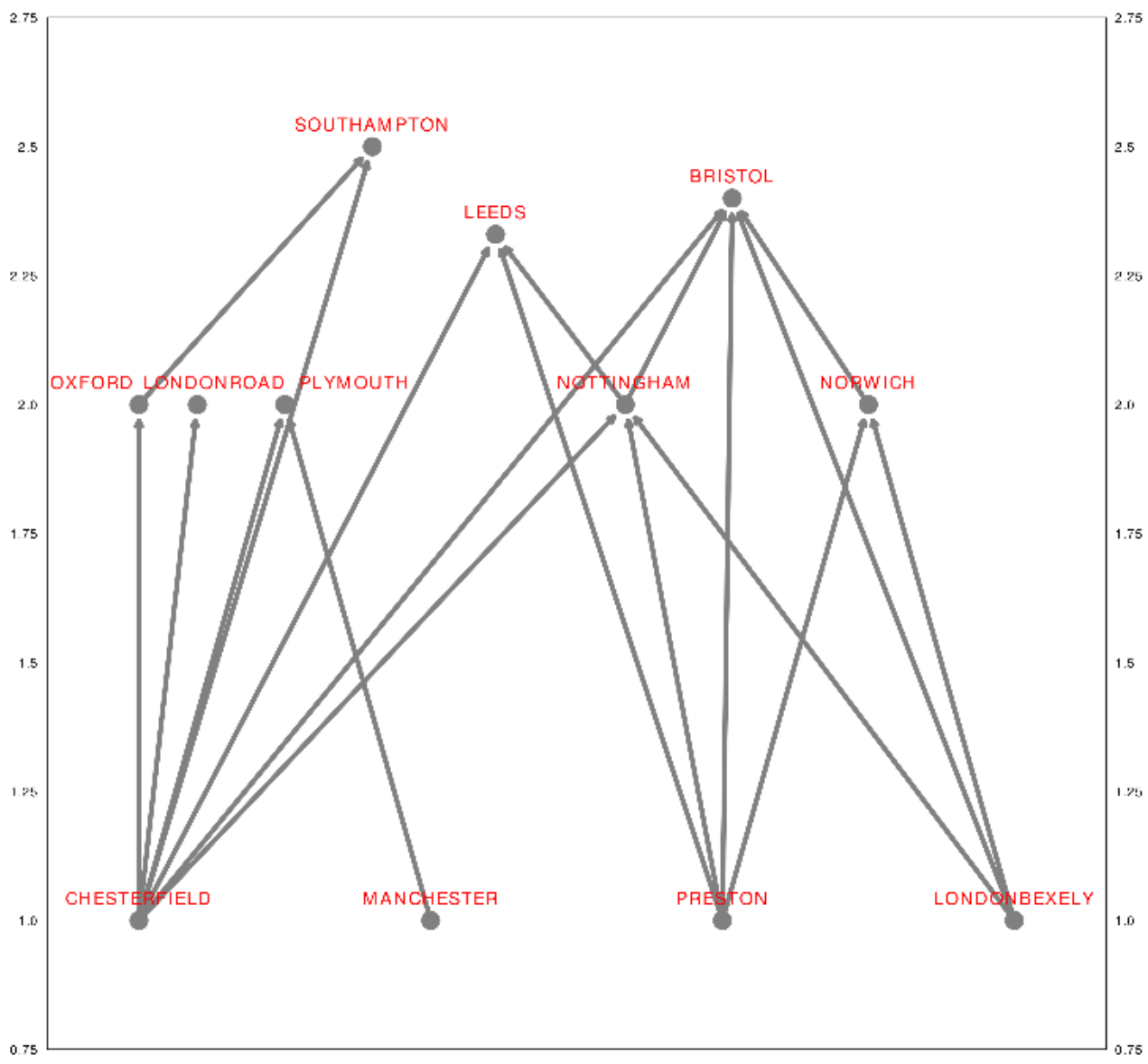

Figure 3. The hierarchical structure and basal nodes of causal network including; A) Spring window, B) Summer window, C) Autumn window, and D) Winter window in 2017-2018, UK.

## 4. Discussion:

### *4.1 The effect of meteorological parameters on network structure*

Based on the previous analysis, this connection (network) indicates that meteorological conditions and diurnal emissions from a wide range of common sources (such as traffic), rather than locally specific sources and events, dominate the relative variations of the concentrations of fine particles over long periods (Gehrig et al., 2003). During wintertime, the meteorology is characterized by frequent inversions, forming an efficient obstacle for the distribution and homogenization of PM. As a result, only tight spatially embedded parts of network (below 100Km with the highest percentage of restored network) could 'withstand' meteorological influences and further parts (above 100Km) started to collapse from a network perspective. In winter time, the plausible reason of connecting the cities out of the initial network (81% of connected cities were out of the initial network with distance above 200 Km) might be higher average seasonal wind speeds (in all studied stations), which probably relates to the balance between greater dilution at higher wind speeds and the shorter transport times at these higher wind speeds, which allows less time for dispersion and deposition of particles over further distances (Harrison et al., 2012).

Indeed, it is well known that changes in meteorological parameters (e.g., wind speed and direction, temperature, and rainfall) can significantly affect $PM_{2.5}$ concentrations and formation mechanisms (AQEG, 2012; Vieno, et al., 2016). In addition to primary sources, secondary sources are dependent on meteorological conditions and the abundance of precursors. Secondary aerosols have a significant contribution in $PM_{2.5}$ concentrations in the UK, where a large proportion transboundary secondary $PM_{2.5}$ transferred from Europe is made of nitrate particles in the form of ammonium nitrate (AQEG, 2012; Vieno, et al., 2016). One plausible reason of connection within a network can be common transboundary sources.

The relationship between $PM_{2.5}$ and wind direction can provide valuable insight into the sources of the measured concentrations. With this in mind, there is a remarkable consistency in the patterns across the Group A and Group B in the UK. There is, however, a subtle difference among cities in the south (Group B) and those in the north or close to northern part (Group A) of the UK (Harrison et al., 2012). High $PM_{2.5}$ concentrations in Group B (southern sites) are more associated with winds from the east through to southeast, which are frequently associated with a blocking high pressure over the Nordic countries, giving rise to an easterly or south-easterly air flow that will transport emissions from eastern Europe, the Netherlands and Belgium, and northern Germany to the southern cities in the UK (Harrison et al., 2012;

Barry and Chorley, 2010). In northern parts of the UK the air arriving from the east to southeast sector will not have passed over these same emission sources.
On the other hand, High $PM_{2.5}$ concentrations in Group A (northern cities or close to northern part) are more significant associated with winds from the northeast through to east, likely to arise when a low pressure runs up the English Channel, drawing air northward across European source areas (to mainly be emission sources of precursors of secondary PM), out into the North Sea, then around the top of the low pressure to reach the northern parts of the UK from a north-easterly direction (Barry and Chorley, 2010).

## 5. Conclusion:

In current study, we use $PM_{2.5}$ concentrations in 14 cities in the UK over 52 weeks to infer an undirected correlation and a directed Granger causality network. We show for both network cases (group A & B), two robust spatial communities divide the UK into the northern and southern city clusters, with greater spatial embedding in spring and summer.
Based on the granger causality test, we infer that $PM_{2.5}$ data of cities with the strongest Cross correlation (having the lowest p-value) can be helpful to predict the future $PM_{2.5}$ values in the network. However, there are of course multiple caveats with this statement, some of which are reflected in our discussions around known influences from meteorological and source variability. We leverage on the directed network to infer stability to disturbances via the trophic coherence parameter, whereby we found that winter had the greatest vulnerability.
As already noted, this connection (network) suggests that meteorological conditions and emissions from regional sources rather than specific local sources and events dominate the relative variations of the urban background $PM_{2.5}$ concentrations (Gehrig et al., 2003) using this sparse network data. We know that PM derived from sources in continental Europe, probably as secondary PM, can have a significant role in affecting $PM_{2.5}$ concentrations in parts of the UK (Harrison et al., 2012). However, our study has some limitations including a short period of time over which the network was analysed. Also, to have a better understanding of network, evaluating a predictive network based $PM_{2.5}$ model using meteorological parameters, and contributions from identified clusters in the UK, would be helpful. This work acts as a demonstrator for the information that can be extracted from an undirected correlation and a directed Granger causality network. Further work is needed, alongside ancillary data that might support the extracted relationships such as source apportionment data and transport activity, for example. The approach might also be better suited to more local networks, such as monitoring stations across a city.

*Code availability*. The code for computing the trophic level of each node in the network, the trophic difference and finally trophic coherence (q) of the network with all scripts needed to reproduce the results in this study is available at https://github.com/elluff/python-TrophicCoherence.

**Acknowledgement:**

This project has received funding from the European Union's Horizon 2020 research and innovation programme Marie Skłodowska-Curie Actions Research and Innovation Staff Exchange (RISE) under grant agreement No. 778360.